\documentstyle[fleqn,12pt]{article}
\begin{document}

\begin{titlepage}
\vspace*{0.3cm}
\baselineskip=24pt

\begin{center}
{\Large \bf        Supercurrent and its Quantum Statistical}\\
{\Large \bf        Properties in Mesoscopic Josephson Junction}\\
{\Large \bf        in the Presence of Nonclassical Light Fields}
\end{center}
\vspace*{0.3cm}

\begin{center}
{\bf            Le-Man Kuang\dag\ddag, Yiwen Wang\dag\ and Mo-Lin Ge\dag} \\
\end{center}

\begin{center}
\dag{\small {\it           Theoretical Physics Division,
                       Nankai Institute of Mathematics,}}\\
{\small {\it           Tianjin, 300071, People's Republic of China}}\\

\ddag{\small {\it       Department of Physics and Institute of Physics, Hunan
Normal University,}}\\
{\small {\it           Hunan 410006, People's Republic of China}}
\end{center}
\vspace*{0.3in}

\begin{abstract}
\baselineskip=24pt

In this paper, we study the supercurrent in a mesoscopic Josephson
junction (MJJ) and its quantum statistical properties in the presence of
nonclassical light fields. We investigate in detail the influence of external
nonclassical light fields on current-voltage step structures of the MJJ. We
also study in detail quantum statistical properties of the supercurrent when
the external quantum electromagnetic fields are even and odd
coherent-state light fields. It is shown that the supercurrent in the MJJ
exhibits both squeezing effect and quantum coherences. It is demonstrated
that the MJJ can feel the difference not only between classical light fields
and nonclassical light fields but also between different nonclassical light
fields.\\

\end{abstract}
\end{titlepage}
\newpage
\section{Introduction}
\baselineskip=24pt

In last few years considerable attention has been devoted to mesoscopic
Josephson junctions (MJJ) [1-6]
due to the prediction of new quantum effects [1,7-10]
in these small-capacitance junctions. Much work concentrated
on the macroscopic quantum tunnelling of the phase of current-biased junctions
and the influence of dissipation on them [11-15].
It is well known that the effect of classical electromagnetic fields(EMF)
on macroscopic Josephson junctions has been well understood.
The Shapiro steps[16] are the result that the external static EMF and
time-varying EMF work together on Josephson junctions.
It is natural to consider the effect of quantum EMF on Josephson junctions.
In particular, it is an interesting topic to study the effect of quantum EMF
on the MJJ since it can respond to the quantum nature of the external EMF.

On the other hand, nonclassical light fields~\cite{r17} have been studied
extensively in the recent years and produced experimentally. These studies
emphasize nonclassical effects of light fields such as squeezing effect and
sub-Poissonian distribution to reflect the quantum nature of light which can
not be understood classically. Even and odd coherent states(CS)\cite{r18,r19}
are two types of important nonclassical states of light and have wide
applications
in physics [21-24]. Although even and odd CS can be decomposed into
superpositions of the Glauber CS, their properties are quite different
from those of the Glauber CS since the former are nonclassical states while the
latter are classical ones\cite{r25}. On the other hand, even CS and odd CS
exhibit significantly different nonclassical characteristic separately
\cite{r19,r26}. Even CS exhibits quadrature squeezing but does not
sub-Poissonian photon statistics while odd CS exhibits sub-Poissonian
but does not quadrature squeezing.

More recently, Vourdas \cite{r27} has studied the effect of external
quantum EMF on the MJJ by the use of a circular superconducting device
with a MJJ in the presence of quantum EMF. He has
investigated the step structures of the current and voltage of the MJJ.
It has been shown that these step structures are similar to Shapiro
steps in macroscopic Josephson junctions, but their details depend on the
nature of quantum EMF.

The aim of this paper is to study systematically properties of the
supercurrent through the MJJ in the presence of nonclassical EMF by taking
even and odd CS as an example of nonclassical EMF and discuss effects of
the external nonclassical EMF on these properties concretely.
We shall study the current-voltage step structures of the MJJ and quantum
statistical properties of the supercurrent. This paper is organized as follows:
in Sec.2 the current-voltage step structures (CVSS)
of supercurrents in the MJJ and the effect of nonclassical EMF on them are
investigated. Sec.3 and Sec.4 are devoted to study quantum
statistical properties of the supercurrent. Squeezing properties and quantum
coherences are investigated in detail for even and odd CS EMF. The results are
summarized and conclusions are drawn in the last section.

\section{CVSS of the MJJ in the presence of even and odd coherent-state EMF}

Consider a MJJ with a capacitance $C$ in a circular superconducting device
\cite{r27}. The MJJ is described by the following Hamiltonian ($\hbar=c=1$)
\begin{equation}
H=\frac{(q+Q)^2}{2C}+E_J(1-\cos \theta) \label{e1}
\end{equation}
where the first term is the charging energy, and the second one is the
Josephson coupling energy; $q$ is the charge on one of the electrodes
of the MJJ, and $Q$ is the external charge;
$\theta$ is the phase difference of the superconducting
order parameters in the two electrodes; and $E_J$ is the Josephson
coupling constant.

Capacitance $C$ of conventional Josephson junctions and the temperatures
where we usually operate allow us to neglect the charging energy in (\ref
{e1}) and to forget the quantum mechanical origin of the Josephson effect.
In stead, we view the Josephson junction as a classical object whose state
is characterized by a sharp value of the phase difference across the
junction. However, In the case of the MJJ,
the classical descriptions is no longer sufficient, and the charging
energy is not negligible. We have to describe the MJJ in a fully quantum
mechanical way. In this case, the charge $q$ and the phase
difference $\theta$ are a pair of quantum mechanical conjugation variables
with $\hat q=-i2e\partial_{\theta}$, they satisfy the canonical commutation
relation:
\begin{equation}
[{\hat \theta},\hat q]=i2e \label{e2}
\end{equation}
The Hamiltonian (\ref{e1}) is then rewritten as
\begin{equation}
H=-\frac 12 E_c(\partial_{\theta}+i\frac Q{2e})^2+E_J(1-\cos{\hat \theta})
\label{e3}
\end{equation}
where the Josephson coupling energy related to the critical current $I_{cr}$
and the charging energy associated with the Cooper-pair charge $2e$ are
given by, respectively,
\begin{equation}
E_J=\frac{I_{cr}}{2e},\ \ E_c=\frac{(2e)^2}{C^2}\label{e4}
\end{equation}

Starting with the Hamiltonian (\ref{e3}), one can find that the
Josephson-current operator $\hat I$  in the Heisenberg picture takes the form:
\begin{equation}
{\hat I}(t)= I_{cr}\sin {\hat \theta}(t)\label{e5}
\end{equation}

As is known, the Shapiro steps on macroscopic Josephson junctions is the
result which the classical static electromotive force work together with the
classical time-varying electromotive force. In the present case,
we still impose the classical static electromotive force $V_0$ but the
quantum time-varying electromotive force on the MJJ through applying a
classical magnetic flux $\phi=V_0 t$ and a ac quantum single-mode magnetic
field of angular frequency $\omega_1$, respectively. Then, the external
voltage induced by the quantum magnetic field in the Heisenberg picture can be
written as
\begin{equation}
V_{ex}(t)=\frac{i\omega_1}{\sqrt{2}}(e^{i\omega_1t}a^{+}-
e^{-i\omega_1t}a)\label{e7}
\end{equation}
where $a$ and $a^{+}$ are the annihilation and creation operators of the
quantum EMF.

we consider the case $\omega_1  \gg \sqrt{E_cE_J}$ where $\sqrt{E_cE_J}$ is the
angular frequency of the MJJ when it is approximated as a harmonic oscillator.
In this case the Hamiltonian of the MJJ itself produces an evolution which is
much slower than the Hamiltonian describing the effect from external fields,
and the phase evolves approximately in the form [27]:
\begin{equation}
e^{i \hat{\theta}(t)} = e^{i\omega_0t}D(\xi e^{i\omega_1t})\label{e8}
\end{equation}
where $\omega_0=2eV_0$, $\xi=\sqrt{2}e$\ ($e$ is the charge of the electron),
and $D(Z)$ is the displacement operator defined by
\begin{equation}
D(Z)=e^{Za^+-Z^*a}
\end{equation}
Then the expectation value of the Josephson current $I$ can be given by
tracing of the current operator $\hat I$ over the density operator of the
external
quantum EMF,
\begin{equation}
I=I_{cr} \mbox{Im}\left\{e^{i\omega_0t}\mbox{Tr}[\rho
D(\xi e^{i\omega_1t})]\right\} \label{e9}
\end{equation}

It is interesting to note that the density operator $\rho$ of the external
quantum EMF controls the operation of the MJJ in the sense that the Josephson
current and also all the other quantities, one might wish to calculate, depend
on
the density operator $\rho$. So that we can change the state in which the
junction operates through changing the density matrix of the external quantum
EMF. This gives a way to realize the controlling of the Josephson tunnelling.

 We now consider the current-voltage step structures of the MJJ
when the external quantum EMF are in even and odd coherent states,
respectively,
and investigate the effect of quantum nature of the external quantum EMF on the
step structures.

The number-state representations of even and odd CS are defined by
\begin{eqnarray}
\mid\!\! Z\rangle _e &=& N_e(\mid\!\! Z\!\mid
)\sum^{\infty}_{n=0}\frac{Z^{2n}}{\sqrt{(2n)!}}
\mid\! 2n\rangle,\ \ N_e(\mid\!\! Z\!\mid )=\cosh ^{-\frac 12}(\mid\!\! Z\!\mid
^2) \label{e10}\\
\mid\!\! Z\rangle _o &=& N_o(\mid\!\! Z\!\mid
)\sum^{\infty}_{n=0}\frac{Z^{2n+1}}{\sqrt{(2n+1)!}}
\mid\! 2n+1\rangle,\ \ N_o(\mid\!\! Z\!\mid )=\sinh ^{-\frac 12}(\mid\!\!
Z\!\mid ^2) \label{e11}
\end{eqnarray}
which can be decomposed into superpositions of Glauber CS $\mid\!\! Z\rangle$
and
$\mid\! -Z\rangle$ in the form
\begin{eqnarray}
\mid\!\! Z\rangle _e=A^{\frac 12}_+(\mid\!\! Z\!\mid )(\mid\!\! Z\rangle
+\mid\! -Z\rangle), &&
A^{-1}_+(\mid\!\! Z\!\mid )=2(1+e^{-2\mid\! Z\!\mid ^2}),\label{e12}\\
\mid\!\! Z\rangle _o=A^{\frac 12}_-(\mid\!\! Z\!\mid )(\mid\!\! Z\rangle
-\mid\! -Z\rangle), &&
A^{-1}_-(\mid\!\! Z\!\mid )=2(1-e^{-2\mid\! Z\!\mid ^2}).\label{e13}
\end{eqnarray}
It should be noted that although even and odd
CS have the above simple
decompositions of Glauber CS, they have quite different quantum nature since
even and odd CS light fields are nonclassical light fields while the Glauber
CS light field is the classical one \cite{r19,r25}.

To get the Josephson current, one needs the following matrix elements of the
displacement operator in the CS representation
\begin{eqnarray}
D_{\pm Z, \pm Z}(\xi e^{i\omega_1t}) &\equiv& \langle\pm Z\!\mid D(\xi
e^{i\omega_1t})
\mid\! \pm Z\rangle =\exp (-\frac 12\xi^2\pm i2u),\label{e14}\\
D_{\pm Z, \mp Z}(\xi e^{i\omega_1t}) &\equiv& \langle\pm Z\!\mid D(\xi
e^{i\omega_1t})
\mid\! \mp Z\rangle =\exp (-\frac 12\xi^2-2\mid\!\! Z\!\mid ^2\pm
2v)\label{e15}
\end{eqnarray}
where
\begin{equation}
u=\xi\mid\!\! Z\!\mid \sin(\omega_1t-\varphi),\ \ v=\xi\mid\!\! Z\!\mid
\cos(\omega_1t-\varphi).\label{e16}
\end{equation}
Here we have taken $Z=\mid\!\! Z\!\mid e^{i\varphi}$.

Making use of Eqs.~(\ref{e14}) and (\ref{e15}), from the expressions (\ref{e8})
and (\ref{e9}) one can obtain the expectation values of the phase-difference
operator in the even and odd CS with the results
\begin{eqnarray}
_e\langle Z\!\mid e^{i\theta(t)}\mid\!\! Z\rangle_e &=& 2A_+(\mid\!\! Z\!\mid
)e^{i\omega_0t}
e^{-\frac 12\xi^2}(\cos 2u+e^{-2\mid\! Z\!\mid ^2}\cosh 2v),\label{e17}\\
_o\langle Z\!\mid e^{i\theta(t)}\mid\!\! Z\rangle_o &=& 2A_-(\mid\!\! Z\!\mid
)e^{i\omega_0t}
e^{-\frac 12\xi^2}(\cos 2u-e^{-2\mid\! Z\!\mid ^2}\cosh 2v)\label{e18}
\end{eqnarray}
and the Josephson current in the presence of even and odd coherent-state EMF,
\begin{eqnarray}
I_e &=& 2I_{cr}A_+(\mid\!\! Z\!\mid )e^{-\frac 12 \xi^2}(\cos 2u
+e^{-2\mid\!Z\!\mid ^2}\cosh 2v)
\sin \omega_0 t, \label{e19}\\
I_o &=& 2I_{cr}A_-(\mid\!\! Z\!\mid )e^{-\frac 12 \xi^2}(\cos 2u
-e^{-2\mid\!Z\!\mid ^2}\cosh 2v)
\sin \omega_0 t \label{e20}
\end{eqnarray}

In order to see the CVSS of the above currents, we
expand the rhs of Eqs.~(\ref{e19}) and (\ref{e20}) as follows:
\begin{eqnarray}
I_e &=& 2I_{cr}A_+(\mid\!\! Z\!\mid)e^{-\frac 12
\xi^2}\sum^{\infty}_{n=-\infty}
[ J_{2n}(2\xi\mid\!\! Z\!\mid )\nonumber\\
&&+ e^{-2\mid\! Z\!\mid^2}I_{2n}(2\xi\mid\!\! Z\!\mid )]\sin[(2n\omega_1+
\omega_0)t-2n\varphi], \label{e21}
\end{eqnarray}
\begin{eqnarray}
I_o &=& 2I_{cr}A_-(\mid\!\! Z\!\mid)e^{-\frac 12
\xi^2}\sum^{\infty}_{n=-\infty}
[ J_{2n}(2\xi\mid\!\! Z\!\mid ) \nonumber\\
&&-e^{-2\mid\! Z\!\mid ^2}I_{2n}(2\xi\mid\!\! Z\!\mid )]\sin[(2n\omega_1+
\omega_0)t-2n\varphi] \label{e22}
\end{eqnarray}
where $J_n(x)$ and $I_n (x)$ are Bessel and modified Bessel functions,
respectively, and we have used the formulae:
\begin{equation}
e^{ix\sin \theta} =\sum^{\infty}_{n=-\infty}J_n(x)e^{in \theta},\ \ \
e^{x\cos \theta} =\sum^{\infty}_{n=-\infty}I_n(x)e^{in \theta}.\label{e23}
\end{equation}
The dc components of the Josephson currents can be obtained through taking
the time meanvalues of the expressions (\ref{e21}) and (\ref{e22}) in the
interval $(0,t)$. It can be seen from Eqs.~(\ref{e21}) and (\ref{e22})
that for even CS and also for odd CS when the dc voltage $V_0$ obeys the
following condition
\begin{equation}
2n\omega_1-\omega_0=0,\ \  \mbox{i.e.,}\ \ V_0=\frac{n\omega_1}e, \label{e24}
\end{equation}
we get Shapiro dc currents similar to those in macroscopic junctions.
Their expressions take the form
\begin{equation}
I^{dc}_e = 2I_{cr}A_+(\mid\!\! Z\!\mid )e^{-\frac 12 \xi^2}\left[
J_{-2n}(2\xi\mid\!\! Z\!\mid )+e^{-2\mid\!Z\!\mid ^2}I_{-2n}(2\xi\mid\!\!
Z\!\mid )\right]
\sin(2n\varphi), \label{e25}
\end{equation}
\begin{equation}
I^{dc}_o = 2I_{cr}A_-(\mid\!\! Z\!\mid )e^{-\frac 12 \xi^2}\left[
J_{-2n}(2\xi\mid\!\! Z\!\mid )-e^{-2\mid\!Z\!\mid ^2}I_{-2n}(2\xi\mid\!\!
Z\!\mid )\right]
\sin(2n\varphi). \label{e26}
\end{equation}

It can be seen from Eqs.~(\ref{e24}), (\ref{e25})
and (\ref{e26}) that even CS and odd CS have similar CVSS. Especially,
they share the same voltage step structures when $\varphi=\frac{k\pi}n\
(n\neq 0,\ k=0,\pm 1,\pm 2,\ldots)$, while the dc components
vanish, i.e., $I^{dc}_e=I^{dc}_o=0$; and when $\mid\!\! Z\!\mid \gg 1$, we find
that even
CS and odd CS have almost equal dc components, that is $I^{dc}_e\doteq I^{dc}
_o\doteq I_{cr} e^{-\frac 12\xi^2}J_{-2n}(2\xi\mid\!\! Z\!\mid
)\sin(2n\varphi)$. Thus,
when $\mid\!\! Z\!\mid \gg 1$, even and odd CS have the same CVSS.
Nevertheless, when
$\mid\!\! Z\!\mid \ll 1$, from Eqs.~(\ref{e25}) and (\ref{e26}) we have
$I^{dc}_o\doteq 0$
and
$
I^{dc}_e\doteq \frac 12 I_{cr}e^{-\frac 12 \xi^2}\left[ J_{-2n}(2\xi\mid\!\!
Z\!\mid )
+I_{-2n}(2\xi\mid\!\! Z\!\mid )\right]\sin (2n\varphi)
$
which indicate that when
$\mid\!\! Z\!\mid \ll 1$ even and odd CS have quite different CVSS. Therefore,
the
MJJ can feel the difference between different external nonclassical EMF.

On the other hand, the CVSS of even and odd CS are also different from that
of the Glauber CS \cite{r27}. The voltage jumps (\ref{e24}) in the former
are double in size in comparison with the voltage jumps in the latter.
This means that the electrostatic energy that the Cooper pair loses during
tunnelling in the case of even and odd CS is twice of that in the Glauber
CS case. As concerns the dc components of the Josephson currents, the dc
components for even and odd CS have only one direction for all steps for
a given $Z$ while for the Glauber CS the dc components have two directions
which correspond to $n$ being even and odd, respectively. Taking into account
the results in Ref.~\cite{r27}, we can find that the Glauber CS, CS with
thermal noise, and CS with partially randomized phase have the same CVSS,
while even and odd CS, squeezed vacuum states exhibit the same CVSS.
This reflects the fact  that the former three types of states are classical
states
while the latter three ones are nonclassical states. Therefore, we can
conclude that the MJJ can also feel the difference between the external
nonclassical EMF and classical ones.

It is worthwhile mentioning that the existence of the CVSS of the MJJ
depends strongly on the form of the density matrix describing the external
quantum EMF. Those states which off-diagonal elements of their density matrices
in the number-state representation are nonzero, such as the Glauber CS,
even and odd CS, squeezed vacuum states, exhibit the CVSS of the MJJ,
while for those states which off-diagonal elements of the density matrices
are zero, such as CS with randomized phase, thermal states and number
states, there do not exist the CVSS. This indicates that the existence of
the CVSS of the MJJ are determined by the external dc voltage $V_0$ and
the off-diagonal elements of the density matrices of the external quantum EMF.
In other words, the step structures depends on the dc voltage $V_0$ and the
quantum
interference among the number states of the external quantum EMF. In this
sense, the step structures reflect to some extend quantum coherences of the
external quantum EMF.

\section{Squeezing properties of supercurrents in the MJJ in the presence
of quantum EMF}

Squeezing is one of the most important concepts in quantum optics
\cite{r25,r28}. It has potential applications in low-noise optical
communications
and high-precision interferometic measurements. It is of importance to study
squeezing behaviors of phase in quantum mechanics \cite{r29,r30}. As is
known, the supercurrent in a Josephson junction is a quantum-phase-correlation
phenomenon. So it is interesting to study squeezing properties of the
supercurrent. In this section, we discuss squeezing of the supercurrent in
the MJJ in the presence of quantum EMF.

Consider two noncommuting Hermitian operators $\hat A$ and $\hat B$ acting on
the state space of the external quantum EMF. The variances (fluctuations)
of these operators $\langle(\Delta\hat A)^2\rangle=\langle \hat{A}^2\rangle
-\langle \hat A \rangle ^2$ and $\langle(\Delta\hat B)^2\rangle
=\langle \hat{B}^2\rangle -\langle \hat B \rangle ^2$ obey the Heisenberg
uncertainty relation
\begin{equation}
\langle(\Delta\hat A)^2\rangle \langle(\Delta\hat B)^2\rangle \geq
\frac 14\mid\! \langle [\hat A, \hat B]\rangle\!\mid ^2 \label{e27}
\end{equation}
which is fundamentally different from that of the position and momentum
operators of a quantum mechanical particle $\langle (\Delta \hat x)^2\rangle
\langle (\Delta \hat p)^2 \rangle \geq \frac 14$ since the commutator
$[\hat A, \hat B]$ on the rhs of equation(\ref{e27}) is not a $C$ number
but an operator which makes the rhs of the uncertainty relation(\ref{e27})
to be state-dependent. Uncertainty relation with a state-dependent rhs is well
known from earlier studies of the atomic coherent states \cite{r31,r32,r33}.
Following W\'odkiewicz and Eberly \cite{r33} we shall say that the variances
of the operators $\hat A$ and $\hat B$ are squeezed if
\begin{equation}
\langle(\Delta\hat A)^2\rangle < \frac 12\mid\!\langle [\hat A, \hat
B]\rangle\!\mid
\ \ \mbox{or}\ \ \langle(\Delta\hat B)^2\rangle
< \frac 12\mid\!\langle [\hat A, \hat B]\rangle\!\mid  \label{e28}
\end{equation}
In order to measure the degree of squeezing of these operators, we
introduce two squeezing parameters $S_A$ and $S_B$ defined by
\begin{equation}
S_A=\frac{\langle(\Delta\hat A)^2\rangle -\frac 12\mid\!\langle [\hat A,
  \hat B]\rangle\!\mid } {\frac 12\mid\!\langle [\hat A, \hat B]\rangle\!\mid
}, \ \
S_B=\frac{\langle(\Delta\hat B)^2\rangle -\frac 12\mid\!\langle [\hat A,
  \hat B] \rangle\!\mid } {\frac 12\mid\!\langle [\hat A, \hat B]\rangle\!\mid
}
  \label{e29}
\end{equation}
The squeezing condition now takes the simple form:
\begin{equation}
S_A<0 \ \ \mbox{or}\ \ S_B<0 \label{e30}
\end{equation}
The maximum(100\%) squeezing of variance $\langle (\Delta \hat X)^2\rangle$
corresponds to $S_X=-1\ (\hat X=\hat A\ \mbox{or}\ \hat B)$.

For the present case of the MJJ in the presence of quantum EMF, from the
canonical commutation relation of the charge on the electrodes and the
phase difference operator (\ref{e2}), we can obtain the commutation
relation between the supercurrent operator and the charge operator,
\begin{equation}
[\hat I, \hat q]=-i 2e I_{\mbox{cr}}\cos \theta \label{e31}
\end{equation}
Then, the degree of the squeezing of the Josephson current is given by
\begin{equation}
S_I=\frac{\langle(\Delta\sin \hat\theta)^2\rangle -e\mid\!\langle
\cos\hat\theta
  \rangle\!\mid } {e\mid\!\langle \cos \hat \theta \rangle\!\mid } \label{e32}
\end{equation}
For simplicity we have taken  $I_{\mbox{cr}}$ as the unit of  measuring
the supercurrent.

For even and odd CS, it is easy to get
\begin{eqnarray}
_e\langle Z\!\mid e^{i{\hat \theta}(t)}\mid\!\! Z\rangle_e &=&
e^{i\omega_0t}\left[ D_{Z,Z}(\xi
e^{i\omega_1 t})+D_{-Z,Z}(\xi e^{i\omega_1 t})\right. \nonumber\\
&&+\left. D_{Z,-Z}(\xi e^{i\omega_1 t})
+D_{-Z,-Z}(\xi e^{i\omega_1 t})\right],\label{e33}\\
_o\langle Z\!\mid e^{i{\hat \theta}(t)}\mid\!\! Z\rangle_o &=&
e^{i\omega_0t}\left[ D_{Z,Z}(\xi
e^{i\omega_1 t})-D_{-Z,Z}(\xi e^{i\omega_1 t})\right. \nonumber\\
&&-\left. D_{Z,-Z}(\xi e^{i\omega_1 t})
+D_{-Z,-Z}(\xi e^{i\omega_1 t})\right]\label{e34}
\end{eqnarray}
where the matrix elements of the displacement operator are given by
Eqs.~(\ref{e14}) and (\ref{e15}).

{}From Eqs.~(\ref{e33}) and (\ref{e34}) we find that
\begin{equation}
\hspace*{-1.0cm}_e\langle Z\!\mid \left( \begin{array}{c}\sin {\hat \theta}\\
\cos \theta \end{array} \right)\mid\!\! Z \rangle_e
= 2A_+(\mid\!\! Z\!\mid )e^{-\frac 12\xi^2}\left[\cos 2u +e^{-2\mid\! Z\!\mid
^2}
\cosh 2v\right] \left( \begin{array}{c}\sin \omega_0t\\ \cos \omega_0t
\end{array}\right)\label{e35}
\end{equation}
\begin{equation}
\hspace*{-1.0cm}
_o\langle Z\!\mid \left( \begin{array}{c}\sin {\hat \theta}\\ \cos \theta
\end{array} \right)\mid\!\! Z \rangle_o = 2A_-(\mid\!\! Z\!\mid )e^{-\frac
12\xi^2}
\left[\cos 2u -e^{-2\mid\! Z\!\mid ^2}
\cosh 2v\right] \left( \begin{array}{c}\sin \omega_0t\\ \cos \omega_0t
\end{array}\right)\label{e36}
\end{equation}
Then, we obtain the fluctuations of the supercurrent in the MJJ for
even and odd CS
\begin{eqnarray}
\langle (\Delta \sin {\hat \theta})^2\rangle_e &=& \frac 12 -B_+\cos
2\omega_0t-C_+^2\sin ^2\omega_0 t,\label{e37}\\
\langle (\Delta \sin {\hat \theta})^2\rangle_o &=& \frac 12 -B_-\cos
2\omega_0t-C_-^2\sin ^2\omega_0 t\label{e38}
\end{eqnarray}
where
\begin{eqnarray*}
B_{\pm}  &=& A_{\pm}(\mid\!\! Z\!\mid )e^{-2\xi^2}(\cos 4u\pm e^{-2\mid\!
Z\!\mid ^2}
\cosh 4v), \\
C_{\pm}  &=& 2A_{\pm}(\mid\!\! Z\!\mid )e^{-\frac 12\xi^2}(\cos 2u\pm
e^{-2\mid\! Z\!\mid ^2}
\cosh 2v).
\end{eqnarray*}

On substituting Eqs.~(\ref{e35})-(\ref{e38}) into Eq.~(\ref{e32}), we
find the squeezing degree of the supercurrents with respect to even and
odd CS, respectively
\begin{eqnarray}
S_I^e(t) &=& \frac{1-2B_+\cos 2\omega_0t-2C_+^2\sin ^2\omega_0t}
{2e\mid\! C_+\cos \omega_0t\!\mid }-1,\label{e40}\\
S_I^o(t) &=& \frac{1-2B_-\cos 2\omega_0t-2C_-^2\sin ^2\omega_0t}
{2e\mid\! C_-\cos \omega_0t\!\mid }-1\label{e41}
\end{eqnarray}
which indicate that when $t\neq \frac{(2n-1)\pi}{2\omega_0} (n=1,2,\ldots)$ and
\begin{equation}
1-2B_{\pm}\cos 2\omega_0 t-2C_{\pm}\sin^2\omega_0t
<2e\mid\! C_{\pm}\cos \omega_0 t\!\mid  \label{e42}
\end{equation}
we have $S_I^e(t)<0$ and $S_I^o(t)<0$ which means that the supercurrent
exhibits squeezing. However, when $t=\frac{(2n-1)\pi}{2\omega_0}
(n=1,2,\ldots)$,
we have $S_I^e(t)=S_I^o(t)=+\infty$. From the uncertainty relation (\ref{e27})
and Eq.~(\ref{e31}) we know that
\begin{equation}
\langle(\Delta\hat I)^2\rangle \langle(\Delta\hat q)^2\rangle\geq 0\label{e43}
\end{equation}
which means that no squeezing of the supercurrent occurs under this
circumstance.

We plot the time evolution of squeezing of the supercurrent in the MJJ
when $\omega_0=2N\omega_1$ for even and odd CS cases in Fig.1 and Fig.2.
As can be seen from Fig.1, in the case of even CS,
the squeezing is weakened with the increase of the value of $\mid\!\! Z\!\mid$
for a given $N$. Nevertheless, the case of odd CS is contrary to the
even CS case.
{}From Fig.2 we see that the
squeezing is strengthened with the increase of the value of $\mid\!\! Z\!
\mid $ for a given $N$, although there does not exist squeezing of the
supercurrent when $Z$ is very small. It is noted that the singular points
of the squeezing parameters appear periodically with the period $\frac{\pi}
{\omega_0}$ which depends only on the external classical voltage bias,
the number of the singular points increases with the increasing of $N$.
Therefore, the angular frequency of the nonclassical EMF does not affect
periodicity of the supercurrent squeezing. Comparing Fig.1 with Fig.2,
we find that the squeezing curve of the odd CS approaches to that of
the even CS with the increasing of $\mid\!\! Z\!\mid$ for a given $N$.

Since even and odd CS are nonclassical states while the Glauber CS are
classical states, their effects on the squeezing of the supercurrent in the
MJJ are quite different. To see this, we write out the squeezing degree of
the supercurrent for Glauber CS,
\begin{equation}
S_I(t)=\frac{(e^{\frac 12 \xi^2}-e^{-\frac 12 \xi^2})\{1+e^{-\xi^2}
\cos [4\xi\mid\!\! Z\!\mid \sin (\omega_1t-\varphi)+2\omega_0 t]\}}
{2e\mid\! \cos [2\xi\mid\!Z\!\mid \sin (\omega_1-\varphi)+\omega_0 t]\!\mid
}-1\label{e44}
\end{equation}
which indicates that when $2\xi\mid\!\! Z\!\mid \sin
(\omega_1t-\varphi)+\omega_0t\neq
(2n-1)\pi/2$ and
$
(e^{\frac 12 \xi^2}-e^{-\frac 12 \xi^2})$\\
$\{1+e^{-\xi^2}
\cos [4\xi\mid\!\! Z\!\mid \sin (\omega_1t-\varphi)+2\omega_0 t]\}
< 2e\mid\! \cos [2\xi\mid\!\! Z\!\mid \sin (\omega_1-\varphi)+\omega_0 t]\!\mid
$
the supercurrent exhibits squeezing (i.e., $S_I(t)<0$). When
$2\xi\mid\!\! Z\!\mid \sin (\omega_1-\varphi)+\omega_0 t=\frac{(2n-1)\pi}2
(n=1,2,\ldots)$, the squeezing disappears due to $S_I(t)=+\infty$
\newpage
\vspace*{20cm}

\begin{center}
{\small FIG.1: Time evolution of squeezing of the supercurrent in the MJJ\\
when the external nonclassical EMF is in even CS for $\varphi=\frac{\pi}2$
and\\
(a) $\mid\!\! Z\!\mid =0.5$, $N=5$; (b) $\mid\!\! Z\!\mid =1$, $N=5$;
(c) $\mid\!\! Z\!\mid =0.5$ $N=1$; (d) $\mid\!\! Z\!\mid =0.1$ $N=1$.}
\end{center}

\vspace*{20cm}

\begin{center}
{\small FIG.2: Time evolution of squeezing of the supercurrent in the MJJ\\
when the external nonclassical EMF is in odd CS for $\varphi=\frac{\pi}2$ and\\
(a) $\mid\!\! Z\!\mid =0.5$, $N=5$; (b) $\mid\!\! Z\!\mid =1$, $N=5$;
(c) $\mid\!\! Z\!\mid =0.5$ $N=1$; (d) $\mid\!\! Z\!\mid =0.1$ $N=1$.}
\end{center}

\newpage

\vspace*{16cm}

\begin{center}
{\small FIG.3: Time evolution of squeezing of the supercurrent in the MJJ\\
when the external quantum EMF is in the Glauber CS for $\varphi=\frac{\pi}2$
and\\
(a) $\mid\!\! Z\!\mid =10$, $N=5$; (b) $\mid\!\! Z\!\mid =10$, $N=1$;
(c) $\mid\!\! Z\!\mid =1$ $N=1$.}
\end{center}

In Fig.3, we plot the time evolution of the squeezing of the supercurrent in
one period in the presence of the Glauber coherent-state EMF when the step
condition is satisfied, i.e., $\omega_0=N\omega_1$. Obviously, the squeezing
of this case is quite different from that of even and odd CS. From (a)
and (b) in Fig.3, we can see that for a given $\mid\!\! Z\!\mid $,
oscillations
of $S_I(t)$ are strenghtened during the preceding half period. From (b) and
(c) we can see that
for a given value of $N$, the oscillatary behaviors are weakened and the
number of the singular points in every period decreases with the
decreasing of $\mid\!\! Z\!\mid$.

\section{Quantum coherences of supercurrent in the MJJ in the presence of
quantum EMF}

In this section we consider another important aspect of statistical
properties of the supercurrent in the MJJ, that is its quantum coherences
which describe the degree of quantum phase correlation in the supercurrent.
The degree of the second-order quantum coherence can be obtained by evaluating
the second-order correlation function of the supercurrent at two different
spacetime points. If purely temporal correlations are of interest, the relevant
correlation function in the normalized form can be defined by
\begin{equation}
g^{(2)}(t,t+\tau)=\frac{\langle :\hat I(t)\hat I(t+\tau) :\rangle}
   {\langle \hat I(t)\rangle \langle \hat I(t+\tau)\rangle} \label{e46}
\end{equation}
where the bracket ``$\langle\ \ \rangle$'' denotes taking expectation values
over states of the external quantum EMF, and ``: :'' stands for normal
ordering of quantum  mechanical operators. Following considerations in
quantum optics, We are interested in behaviors of the
second-order correlation function (\ref{e46}) at $\tau=0$. Then, from
Eq.~(\ref{e46}) we get
\begin{equation}
g^{(2)}(t)\equiv g^{(2)}(t,t)=\frac{\langle :{\hat I}^2(t) :\rangle}
   {\langle \hat I(t)\rangle ^2} \label{e47}
\end{equation}
Substituting the expression of the supercurrent operator (\ref{e5})
into the above equation and making use of the formula
\begin{equation}
: D(2\xi e^{i\omega_1 t}): =e^{2\xi^2}D(2\xi e^{i\omega_1 t})\label{e48}
\end{equation}
we find the general expression of the second-order correlation function
in the following form
\begin{equation}
g^{(2)}(t)=\frac{1-e^{2\xi^2}\mbox{Re}\langle e^{i2\theta}\rangle}
{2\langle \sin \theta \rangle^2}\label{e49}
\end{equation}
For even and odd CS, we find that
\begin{equation}
g^{(2)}_e=\frac{1-2B_+\cos 2\omega_0 t}{2C_+\sin^2\omega_0t},\ \
g^{(2)}_o=\frac{1-2B_-\cos 2\omega_0 t}{2C_-\sin^2\omega_0t}\label{e50}
\end{equation}
where $B_{\pm}$ and $C_{\pm}$ are given by Eq.~(\ref{e41}).

{}From Eq.~(\ref{e50}), it can be seen that when $t=\frac{n\pi}{\omega_0}
 (n=0,1,2,\ldots)$ we have $g^{(2)}_e(t)=g^{(2)}_o(t)=+\infty$ which indicates
that quantum coherences vanish since the mean value of the supercurrent is zero
at these points.

\vspace*{11cm}
\begin{center}
{\small FIG.4: Time evolution of quantum coherence of the supercurrent \\
when the external nonclassical EMF is in even CS for $\varphi=\frac{\pi}2$\\
and (a) $\mid\!\! Z\!\mid =0.5$, $N=1$; (b) $\mid\!\! Z\!\mid =0.5$, $N=2$.}
\end{center}

In Fig.4 and Fig.5 we plot the time evolution of quantum coherences of the
supercurrent in the MJJ when $\omega_0=2N\omega_1$ for even CS and odd CS,
respectively. As can be seen from these figures that when the external
nonclassical EMF is in even and odd CS, the supercurrent in the MJJ
exhibits similar quantum coherences during the time evolution.

\vspace*{10.5cm}
\begin{center}
{\small FIG.5: Time evolution of quantum coherence of the supercurrent \\
when the external nonclassical EMF is in odd CS for $\varphi=\frac{\pi}2$\\
and (a) $\mid\!\! Z\!\mid =0.5$, $N=1$; (b) $\mid\!\! Z\!\mid =0.5$, $N=2$.}
\end{center}

It is worthwhile mentioning that when the external quantum EMF is in Glauber
CS, the time evolution of quantum coherences of the supercurrent in the MJJ
is quite different from that of the even and odd CS cases. For the case of
the Glauber CS, it is quite easy to find the second order correlation
function to be $g^{(2)}(t) =e^{\xi}\doteq 1$ which means that the quantum
coherence of the supercurrent in the presence of the coherent-state EMF
does not change during the time evolution. Therefore, nonclassical states
and classical states of the external quantum EMF lead to quite different
time evolution of quantum coherences of the supercurrent in the MJJ,
in this sense, the MJJ can distinguish the external nonclassical light
fields and classical light fields.

\section{Concluding remarks}

We have studied some properties of the supercurrent in the MJJ in the
presence of nonclassical light fields, the CVSS, squeezing effect and
quantum coherences. In particular, we have investigated in detail these
properties when the external quantum  EMF is in even and odd CS. We have
found that the CVSS of the MJJ depends strongly on the off-diagonal elements
of the density matrix which describes the external quantum EMF.
We have shown that the supercurrent in the MJJ exhibits both squeezing and
quantum coherences. However, when the CVSS of the MJJ is satisfied,
squeezing and quantum coherences vanish periodically due to the periodical
appearance of singular points of the squeezing parameter and the second-order
correlation function of the supercurrent. we have also shown that the MJJ can
feel
the difference not only between classical light fields and nonclassical
light fields but also betwwen different nonclassical light fields.
This indicates that the MJJ are very sensitive devices that
can respond to the quantum nature of external nonclassical light fields.
Therefore, the MJJ is a sensitive detector to nonclassical light fields.
We believe that the results obtained in this paper could be useful in the
study of the controlling of Josephson tunnelling and the detection of
nonclassical light fields in quantum optics.\\

The research was supported in part by the National Nature Science
Foundation of China.


\newpage
\begin{thebibliography}{99}
\bibitem{r1}G. Schon and A. D. Zaikin, {\em Phys. Rep.} {\bf 198}, 237(1990).

\bibitem{r2}D. V. Averin and K. K. Likharev, {\em J. Low Temp. Phys.}
{\bf 62}, 345(1986).

\bibitem{r3}A. Widom, G. Megaloudis, T. D. Clark and R. J. Prance,
{\em J. Phys.} {\bf A15}, 1561(1982).

\bibitem{r4}A. Widom, G. Megaloudis, T. D. Clark, J. E. Mutton, R. J. Prance
and H. Prance,
{\em J. Low Temp. Phys.} {\bf 57}, 651(1984).

\bibitem{r5}F. W. J. Hekking, L. I. Glazman, K. A. Matreev and R. I. Shekhter,
{\em Phys. Rev. Lett.} {\bf 70}, 4138(1993).

\bibitem{r6}G. Ingold and H. Grabert, {\em Phys. Rev.} {\bf B50}, 395(1994);
R. Bauerrschmitt, J. Siewert, Y. V. Nazarov and A. A. Odintsov,
{\em Phys. Rev.} {\bf B49}, 4076(1994).

\bibitem{r7}K. K. Likharev and A. B. Zorin, ``Proceedings of the International
Conference on Low Temperature Physics -LT-17'', edited by U. Eckern, A. Schmid,
W. Weber and H. W\"uhl (Elsevier, Amsterdam, 1984), p.1153.

\bibitem{r8}K. K. Likharev and A. B. Zorin, {\em J. Low Temp. Phys.}
{\bf 59}, 347(1985).

\bibitem{r9}D. V. Averin, A. B. Zorin and K. K. Likharev,
{\em Sov. Phys. -JETP} {\bf 61}, 407(1985).

\bibitem{r10}M. B\"uttiker, {\em Phys. Rev.} {\bf B36}, 3548(1987).

\bibitem{r11}A. O. Caldeira and A. J. Leggett, {\em Phys. Rev. Lett.}
{\bf 46}, 211(1981);
{\em Ann. Phys.}(NY) {\bf 149}, 374(1983).

\bibitem{r12}R. F. Voss and R. A. Webb, {\em Phys. Rev. Lett.} {\bf 47},
265(1981); S. Washburn, R. A. Webb, R. F. Voss and S. M. Fairs,
{\em Phys. Rev. Lett.} {\bf 54}, 2712(1985).

\bibitem{r13}L. D. Jackel, J. P. Gordon, E. L. Hu, R. E. Howard, L. A. Fetter,
D. M. Tennant, R. W. Epworth and J. Kurkij\"arvi, {\em Phys. Rev. Lett.}
{\bf 47}, 697(1981).

\bibitem{r14}M.H. Devoret, J.M. Martinis and J. Clarke, {\em Phys. Rev. Lett.}
{\bf 55}, 1908(1985).

\bibitem{r15}D. B. Schartz, B. Sen, C. N. Archie and J. E. Lukens,
{\em Phys. Rev. Lett.} {\bf 55}, 1547(1985).

\bibitem{r16}S. Shapiro, {\em Phys. Rev. Lett.} {\bf 11}, 80(1963).

\bibitem{r17}R. Loudon and P. L. Knight, {\em J. Mod. Opt.}
{\bf 34}, 709(1987);
M. C. Teich and B. E. A. Saleh, {\em Quantum Opt.} {\bf 1}, 153(1989).

\bibitem{r18}J. Perina, ``Quantum statistics of linear and nonlinear
optical phenomena''(Reidel, Dordrecht, 1984), p.78.

\bibitem{r19}V. Bu\u{z}ek, A. Vidiella-Barranco and P. L. Knight,
{\em Phys. Rev.} {\bf A45}, 6570(1992).

\bibitem{r20}N. A. Ansari, L. D. Fiore, M. A. Man'ko, V. I. Man'ko, S. Solimeno
and F. Zaccarria, {\em Phys. Rev.} {\bf A49}, 2151(1994);
A. Vidiella-Barranco, H. Moya-Cessa and V. Bu\u{z}ek,
{\em J. Mod. Opt.} {\bf 39}, 1441(1992).

\bibitem{r21}J. Gea-Banacloche, {\em Phys. Rev.} {\bf A44}, 5913(1991);
V. Bu\u{z}ek, H. Moya-Cessa, P.L. Knight, and S. J. D. Phoenix,
{\em Phys. Rev.} {A45}, 8190(1992).

\bibitem{r22}M. Brune, S. Haroche, J. M. Raimond, L. Davidovich and N. Zagury,
{\em Phys. Rev.} {\bf A45}, 5193(1992).

\bibitem{r23}C. C. Gerry, {\em Opt. Commun.} {\bf 91}, 247(1992);
C. C. Gerry and E. E. Hach III, {\em Phys. Lett.} {\bf A174}, 185(1993).

\bibitem{r24}F. B. Wang and L. M. Kuang, {\em Phys. Lett.} {\bf A169},
225(1992);
L. M. Kuang and F. B. Wang, {\em Phys. Lett.} {\bf A173}, 221(1993).

\bibitem{r25}D.F. Walls, {\em Nature}, {\bf 306}, 141(1983);
P. Tombesi and E. R. Pike, ``Squeezed and non-classical light'' (Plenum,
New York, 1989).

\bibitem{r26}F. B. Wang and L. M. Kuang, {\em J. Phys.} {\bf A26}, 293(1993);
Y. Xia and G. Guo, {\em Phys. Lett.} {\bf A136}, 281(1989).

\bibitem{r27}A. Vourdas, {\em Phys. Rev.} {\bf B49},12040(1994).

\bibitem{r28}For a review on the subject see
H. J. Kimble and D. F. Walls, {\em J. Opt. Soc. Am.} {\bf B4}(1987)(special
issue
on squeezed states of the electromagnetic field);
L. M. Kuang, F. B. Wang
and Y. G. Zhuo, {\em J. Mod. Opt.} {\bf 41}, 1307(1994).

\bibitem{r29}P. Carruthers and M. M. Nieto, {\em Phys. Rev. Lett.} {\bf 14},
387(1965).

\bibitem{r30}L. M.  Kuang and X. Chen, {\em Phys. Rev.} {\bf A50} 4228(1994);
{\em Phys. Lett.} {\bf A186}, 8(1994).

\bibitem{r31}A. Perelomov, ``Generalized coherent states and their
applications'' (Springer-Verlag, Berlin, 1986).

\bibitem{r32}J. M. Radcliffe, {\em J. Phys.}(Paris) {\bf A4}, 313(1971);\\
F. T. Arecchi, E. Courtens, R. Gilmore and H. Thomas,
{\em Phys. Rev.} {\bf A6}, 2211 (1972).

\bibitem{r33}K. W\'odkiewicz and J.H. Eberly, {\em J. Opt. Soc. Am.}
{\bf B2}, 458(1985).
\end{thebibliography}
\end{document}